\documentclass[12pt]{iopart}
\usepackage{iopams}  
\usepackage{graphicx}
\usepackage{bm}
\usepackage{txfonts}

\begin{document}

\title[]{Magnetic field induced charge order in cuprate superconductors: an explanation by spin-vortex-induced loop currents}

\author{Daichi Manabe, Hiroyasu Koizumi}

\address{
Center for Computational Sciences, University of Tsukuba, Tsukuba, Japan
}
\ead{koizumi.hiroyasu.fn@u.tsukuba.ac.jp}
\vspace{10pt}

\begin{abstract}
We present a possible explanation for a recently observed magnetic field induced charge order in cuprate superconductors (Edkins et al.~arXiv:1802.04673 [cond-mat.supr-con]).
We argue that it arises from the reorganization of spin-vortex-induced loop current (SVILC) pattern due to supercurrent-flow caused by the magnetic field. The reorganization is from the 
most stable tiling of $4a \times 6a$ spin-vortex quartets ($a$ is the lattice constant in the CuO$_2$ plane, and a spin-vortex quartet is stable of unit of spin-vortices that contains four holes, four spin-vortices, and four SVILCs) to that of $4a \times 8a$ spin-vortex quartets.
The consequence of this reorganization will lead to the enhancement of $8a$ charge order, and  reduction of $6a$ charge order.
The former is observed in the experiment, but the latter is not confirmed, so far. However, it may be confirmed if the experimental result is carefully reexamined. 
\end{abstract}

%
%
%
%
%

\section{Introduction}

It is now widely-recognized that the high temperature superconductivity in cuprates  (cuprate superconductivity) cannot be explained by the BCS theory.
In order to explain it, a marked departure form the BCS theory is required.

 {\em Spin-vortex-induced loop current theory for superconductivity} is a theory developed to explain the cuprate superconductivity \cite{Koizumi2011,Hidekata2011,HKoizumi2013,HKoizumi2014}. It is also speculated to be enlarged to explain the ordinary superconductivity whose superconducting transition temperature is explained by the BCS theory as well \cite{Koizumi2015b}.
 It explains the persistent current that flows through superconductors, i.e., a macroscopic current flow through the system without voltage drop. It also explains the flux quantization in the units ${ h \over {2e}}$ and the quantization of voltage across the Josephson junction (so-called " Shapiro steps") in the units ${ {hf} \over {2e}}$, where $f$ is the frequency of the radiation field present.
 
 The crucial ingredient of this theory is spin-twisting itinerant motion of electrons; if this occurs the wave function obtained by the requirement of energy minimization becomes a multi-valued function of electron coordinates due to the appearance of spin-vortices created by itinerant electrons.
It is expected that such electron motion occurs in the cuprate, as explained, below.
In hole-doped cuprate superconductors, bulk-sensitive experiments indicate small polaron (pseudo Jahn-Teller small polaron) formation due to strong hole-lattice interaction \cite{Bianconi,Miyaki2008}. This small polaron formation is suppressed in the surface region where an energy gap with d-wave pairing profile is observed, due to the absence of the charge layer that covers the CuO$_2$ plane and stabilizes the polaron. 
The small polarons in CuO$_2$ planes in the bulk give rise to the following two important effects.

\

1.  Appearance of an antiferromagnetic exchange interaction between electrons across the hole occupied sites in the CuO$_2$ plane. 

2.  Appearance of an internal electric field with the component perpendicular to the CuO$_2$ plane around the hole occupied sites in the CuO$_2$ plane.

\

The first effect creates a frustration in spins since another well-known antiferromagnetic exchange interaction exists between electrons in the nearby copper sites;
this interaction is responsible for the antiferromagnetic order in the parent compound.
Due to the competition between the two antiferromagnetic exchange interactions, spin-vortices are created around the hole occupied sites.

The second effect creates a Rashba spin-orbit interaction when electric current exists; actually, the spin-vortices mentioned above creates a Dirac string with $\pi$ flux (singularities of wave functions with sign-change Berry phase around it) at the centers of spin-twisting and it creates the required electric current. Since the electric current is in the CuO$_2$ plane and the internal electric field has a component perpendicular to it, the spin vortices with twisting components in the CuO$_2$ plane should arise.
Over all, spin-vortices and loop currents (we call them, {\em spin-vortex-induced loop currents (SVILCs)}) are created around the hole occupied sites.
A macroscopic supercurrent is generated as a collection of the SVILCs.

In the present work, we first show that the system with SVILCs can have stable current carrying states under nonzero external current feeding conditions.
We may identify that this state is the superconducting state with supercurrent flowing through it.

Next we consider the charge order in the cuprate. 
One of the important issues on the pseudogap phase is the origin of the charge order detected by bulk sensitive experiments, such as nuclear magnetic resonance (NMR) \cite{Wu2011,Wu2013,Wu2015}, and resonant inelastic X-ray scattering (RIXS) \cite{Ghiringhelli2012,Achkar2012,Blackburn2013}.
A remarkable point on this charge order is that it is enhanced by the application of magnetic fields \cite{Wu2011,Chang2012,Sebastian2012}.
Recently, a clear magnetic-field induced charge order is observed by STM  in the cuprate vortex halo \cite{Seamus2018}.
We will argue that it is a consequence of reorganization of spin-vortex induced loop currents by the current flowing.

\section{Essence of SVILC theory of superconductivity}

We explain the essence of SVILC theory of superconductivity in this section. This theory explains the stability of current carrying state. Note that such a stability has never been proved by any theories derived from the BCS one; thus, the SVILC theory is the only microscopic theory so far that can calculate supercurrent flowing stable states.

The wave function of superconducting state is given by
\begin{eqnarray}
{ \Psi({\bf r}_1}, \cdots, {\bf r}_N, t)=\Psi_0({\bf r}_1, \cdots, {\bf r}_N) e^{-{i \over 2} \sum_{\alpha=1}^N \chi ({\bf r}_{\alpha}) }
\end{eqnarray}
where ${\bf r}_i$ is the coordinate of the $i$th electron and $N$ is the number of electrons. $\Psi_0$ is the wave function obtained by energy minimization. It is multi-valued with respect to electron coordinates when spin-twisting itinerant motion of electrons occurs. The phase factor
$e^{-{i \over 2} \sum_{\alpha=1}^{N} \chi ({\bf r}_{\alpha})}$ compensates the multi-valuedness of $\Psi_0$  to make $\Psi$ single-valued, where $\chi$ is an angular variable of period $2\pi$.

For the two-dimensional lattice model for the CuO$_2$ plane for the cuprate, single-particle wave functions are given by
 \begin{eqnarray}
 |\gamma \rangle& =&\sum_{j} e^{\!-\!i { {\chi_j } \over 2}} [e^{-i{{\xi_j} \over 2}}D_{j \uparrow}^{\gamma} {c}^{\dagger}_{j \uparrow}\!+\!e^{i{{\xi_j} \over 2}}D_{j \downarrow}^{\gamma} {c}^{\dagger}_{j \downarrow}] | {\rm vac} \rangle
 \label{gamma}
 \end{eqnarray}
 where $j$ indicates the $j$th site of the lattice describing the position of copper atom, and ${c}^{\dagger}_{j \sigma}$ is the creation operator for electron with spin $\sigma$ at the $j$th site; $\xi_j$ is the angle of spin in the CuO$_2$ plane at the $j$th site,
 and $\chi_j$ is the value of $\chi$ at the $j$th site; parameters $D_{j \uparrow}^{\gamma}$ and $D_{j \downarrow}^{\gamma}$ are obtained from the Hartree-Fock calculation \cite{HKoizumi2013,HKoizumi2014,Koizumi2017}.  The total wave function $\Psi$ is constructed as a sum of Slater determinants using $\{ |\gamma \rangle \}$ as a single-particle wave function basis.
 Actually, calculations below will be done employing a single Slater determinant wave function composed of the lowest energy $N$ single-particle states as the many-body wave function.
 
The multi-valuedness in $\Psi_0$ arises from the phase factors $e^{\pm i{{\xi_j} \over 2}}$ in Eq.~(\ref{gamma}); when $\xi$ is transported along a loop $C_{\ell}$  in the lattice, it shifts as
\begin{eqnarray}
\xi_j \rightarrow \xi_j + 2 \pi w_{\ell}[\xi]
\end{eqnarray}
where $w_{\ell}[\xi]$ is the wining number of function $\xi$ for $C_{\ell}$ given by
\begin{eqnarray}
w_{\ell}[\xi]={1 \over {2 \pi}} \oint_{C_{\ell}} \nabla \xi \cdot d{\bf r}
\end{eqnarray}
which is written for the lattice system as
\begin{eqnarray}
w_{\ell}[\xi]={1 \over {2 \pi}} \sum^{n_{\ell}}_{j=1} \left(\xi_{\ell[j+1] } - \xi_{\ell[j]} \right)
\end{eqnarray}
where $\ell[j]$ is the $j$th site of $C_{\ell}$ with $\ell[n_{\ell}+1]=\ell[1]$, and $n_{\ell}$ is the number of sites in  $C_{\ell}$.

If $w_{\ell}[\xi]$ is odd, it causes the sign-change in $e^{\pm i{{\xi_j} \over 2}}$; this effect may be viewed as the existence of a Dirac string with flux $\pi$ at the centers of spin-vortices

The phase factor $e^{- i{{\chi_j} \over 2}}$ in Eq.~(\ref{gamma}) restores the single-valuedness of the wave function by compensating the sign-change in $e^{\pm i{{\xi_j} \over 2}}$. For that purpose, we impose the following conditions
\begin{eqnarray}
w_{\ell}[\xi]+ w_{\ell}[\chi] = \mbox{Even number }  \mbox{ for all $C_{\ell}$}
\label{constraint}
\end{eqnarray}
With above conditions,
the phase factors $e^{\!-\!i { {\chi_j } \over 2}} e^{\pm i{{\xi_j} \over 2}}$ in $|\gamma \rangle$ become single-valued; thus, the total wave function also becomes single-valued.

In order to impose the condition in Eq.~(\ref{constraint}), we use the method of Lagrange multipliers by considering the following functional
\begin{eqnarray}
F[\nabla \chi ]=E[\nabla \chi]+\sum_{\ell=1}^{N_{\rm loop}} { {\lambda_{\ell}}}\left(  \oint_{C_\ell} \nabla \chi \cdot d {\bf r}-2 \pi w_{\ell} \right)
\label{FunctionalF}
\end{eqnarray}
where $\lambda_{\ell}$'s are Lagrange multipliers, $\{ C_1, \cdots, C_{N_{\rm loop}} \}$ is the basis of loops (i.e., any loop can be constructed from them), and $E[\nabla \chi]$ is the energy functional given by
\begin{eqnarray}
E[\nabla \chi]= \langle \Psi |H[{\bf A}^{\rm em}]| \Psi \rangle=\langle \Psi_{0} |H \left[{\bf A}^{\rm em}-{{\hbar} \over {2e}}\nabla \chi \right]| \Psi_{0} \rangle
\label{energy}
\end{eqnarray}
with ${\bf A}^{\rm em}$ being the electromagnetic vector potential. $w_{\ell}$'s are parameters that impose the condition in Eq.~(\ref{constraint}). Depending on the set of parameters $\{ w_{\ell} \}$, the current flow pattern of SVILCs changes. Thus, a variety of current patters are possible by the choice of $\{ w_{\ell} \}$.

Actually, $\chi$ is evaluated through $\nabla \chi$ obtained as a solution of the following system of equations;
\begin{eqnarray}
{{\delta F[\nabla \chi]} \over {\delta \nabla \chi}}&=&{{\delta E[\nabla \chi]} \over {\delta \nabla \chi}}+\sum_{\ell=1}^{N_{\rm loop}} { {\lambda_{\ell}}} {{\delta } \over {\delta \nabla \chi}} \oint_{C_\ell} \nabla \chi \cdot d {\bf r}=0
\label{Feq1}
\\
 \oint_{C_\ell} \nabla \chi \cdot d {\bf r}&=&2 \pi w_{\ell} 
 \label{Feq2}
\end{eqnarray}

Since $\chi$ and $\xi$ are generally path-dependent multi-valued functions, evaluation of their values at lattice sites must be carefully done. For that purpose, we introduce cuts of bonds in the lattice so that there is only one path $P_{1 \rightarrow k}$ that connects the first site and $k$th site through the bonds in the lattice without crossing any cuts (of bonds); namely, $\chi_k$ and $\xi_k$ are given uniquely as 
\begin{eqnarray}
\chi_k &=&\chi_1 + \int_{P_{1 \rightarrow k}} \nabla \chi \cdot d{\bf r}
\\
\xi_k &=&\xi_1 + \int_{P_{1 \rightarrow k}} \nabla \xi \cdot d{\bf r}
\end{eqnarray}
The jump of values across the cuts  are $2\pi$ multiple of integer for both $\chi$ and $\xi$; thus, multi-valuedness dose not occur in  $e^{\!-\!i { {\chi_j } \over 2}} e^{\pm i{{\xi_j} \over 2}}$ of $|\gamma \rangle$, making $|\gamma \rangle$ a single-valued.

The total energy does not  depend on the value of $\chi_1$ since it just affects the overall constant phase of the wave function.
However, the total energy depends on the value of $\xi_1$ when Rashba interaction exists. 
As will be shown later, this $\xi_1$ dependence is crucial for supercurrent generation under external current feeding.

The external current boundary condition is imposed by adding external loops to functional $F$ as
\begin{eqnarray}
F[\nabla \chi ]=E[\nabla \chi]+\sum_{\ell=1}^{N_{\rm loop}} { {\lambda_{\ell}}}\left(  \oint_{C_\ell} \nabla \chi \cdot d {\bf r}-2 \pi w_{\ell} \right)
+\sum_{\ell=1}^{N^{\rm EX}_{\rm loop}} { {\lambda^{\rm EX}_{\ell}}} \oint_{C^{\rm EX}_\ell} \nabla \chi \cdot d {\bf r}
\label{FunctionalF2}
\end{eqnarray}
where $C^{\rm EX}_{\ell}$ is a external loop that connects a site in the lattice system to another site in the lattice; one of them is the site for flow-in and the other is for flow-out of the external current. Note that $\lambda^{\rm EX}_{\ell}$ is not a Lagrange multiplier; it is determined by the direction and magnitude of the external current through $C^{\rm EX}_\ell$ as boundary conditions.

For the case with external loops, one of the system of equations in Eq.~(\ref{Feq1}) becomes
\begin{eqnarray}
{{\delta E[\nabla \chi]} \over {\delta \nabla \chi}}+\sum_{\ell=1}^{N_{\rm loop}} { {\lambda_{\ell}}} {{\delta } \over {\delta \nabla \chi}} \oint_{C_\ell} \nabla \chi \cdot d {\bf r}+\sum_{\ell=1}^{N^{\rm EX}_{\rm loop}} { {\lambda^{\rm EX}_{\ell}}} {{\delta } \over {\delta \nabla \chi}} \oint_{C^{\rm EX}_\ell}\nabla \chi \cdot d {\bf r}=0
\label{Feq1b}
\end{eqnarray}

Recently, we have developed a method for obtain $\nabla \chi$ without using the above equations. In this method, the fact that the current density is given by
\begin{eqnarray}
{\bf j}=- {{\delta E} \over {\delta {\bf A}^{\rm em}}}={{2e} \over \hbar}  {{\delta E} \over {\delta \nabla \chi}}
\end{eqnarray}
is used. 

For the lattice system, it is given by
\begin{eqnarray}
J_{ j \leftarrow i}= {{2e} \over \hbar}  {{\partial E} \over {\partial \tau_{ j \leftarrow i}}}; \quad  \tau_{ j \leftarrow i}=\chi_j -\chi_i   
\end{eqnarray}
where $J_{ j \leftarrow i}$ is the current through the bond between sites $i$ and $j$ in the direction  $j \leftarrow i$.

Then, the current conservation at site $j$ is given by
\begin{eqnarray}
0=J^{\rm EX} _{ j} + \sum_i {{2e} \over \hbar}  {{\partial E} \over {\partial \tau_{ j \leftarrow i}}}
\label{Feq3}
\end{eqnarray}
where $J^{\rm EX} _{ j}$ is the external current that enters through site $j$. These equations replace those in Eq.~(\ref{Feq1b}).

By employing the above equations, the introduction of external loops in the calculation is not necessary when imposing current feeding boundary conditions.
It also facilitates calculations including the Rashba interaction.

For the lattice system, the number of $\tau_{ j \leftarrow i}$ to be evaluated is equal to the number of bonds. The number of equations in Eq.~(\ref{Feq2}) is equal to the number of plaques of the lattice. 

By imposing the conservation of the current at all sites except one, we have
\begin{eqnarray}
(\mbox{The number of bonds})=(\mbox{The number of plaques})+(\mbox{The number of sites}-1)
\nonumber
\\
\label{Euler1}
\end{eqnarray}

This corresponds to Euler's theorem for the two-dimensional lattice 
\begin{eqnarray}
(\mbox{The number of edges})=(\mbox{The number of faces})+(\mbox{The number of vertices}-1)
\nonumber
\\
\label{Euler2}
\end{eqnarray}

The reason for the minus one in ($\mbox{The number of sites}-1$) in Eq.~(\ref{Euler1}) is due to the fact that the conservation of the total charge is maintained in the calculation; thus, requiring the conservation of current for all sites is redundant by one.  It is interesting that the mathematical expression in Eq.~(\ref{Euler2})
can be interpreted in a physical way as those for the number of unknows and the number of equations in Eqs.~(\ref{Feq2}) and (\ref{Feq3}).

\section{Supercurrent generation by external current feeding}

According to the Dirac equation, the spin-orbit interaction is given by
\begin{eqnarray}
H^{\rm Dirac}_{\rm so}= -{{e \hbar} \over {4 m^2 c^2 }} {\bm \sigma} \cdot \left[ {\bf E}^{\rm em} \times ( {\bf p}+e {\bf A}^{\rm em}) \right]
\end{eqnarray}
where $m$ is the electron mass, ${\bm \sigma}$ the vector of Pauli matrices, ${\bf E}^{\rm em}$ is electric field, ${\bf p}$ the momentum.

In the present model,  the dominant component of ${\bf E}^{\rm em}$ is assumed to be perpendicular to the CuO$_2$ plane ($z$ direction). By neglecting the effect of the magnetic field, we adopt ${\bf A}^{\rm em}=0$. For the case where the SVILC with winding number $+1$ ($-1$), the current flows counterclockwise (clockwise) around the center, thus, the expectation value of ${\bf p}$ exhibits a counter-clockwise (clockwise) circular flow. 
Depending on the spin direction and loop current direction, the value of the above term changes .

We anticipate the spin is polarized in the CuO$_2$ plane ($xy$ plane) due to a Rashba type spin-orbit interaction appearing around the holes.
Thus, we adopt the following Rashba interaction Hamiltonian \cite{Rashba2013} as a model Hamiltonian for the spin-interaction, 
\begin{eqnarray}
H_{\rm so} \!&=&\! \lambda\sum_{h} \Big[ c^{\dagger}_{h\!+\!y \downarrow}c_{h\!-\!x \uparrow}\!-\!c^{\dagger}_{h\!+\!y \uparrow}c_{h\!-\!x \downarrow}\!+\!i(c^{\dagger}_{h\!+\!y\downarrow}c_{h\!-\!x\uparrow}\!+\!c^{\dagger}_{h\!+\!y\uparrow} c_{h\!-\!x\downarrow})
\nonumber
\\
\!&+&\! c^{\dagger}_{h\!+\!x \downarrow}c_{h\!-\!y \uparrow}\!-\!c^{\dagger}_{h\!+\!x \uparrow}c_{h\!-\!y \downarrow}\!+\!i(c^{\dagger}_{h\!+\!x\downarrow}c_{h\!-\!y\uparrow}\!+\!c^{\dagger}_{h\!+\!x\uparrow} c_{h\!-\!y\downarrow})
\nonumber
\\
\!&+&\!c^{\dagger}_{h-x\downarrow}c_{h-y\uparrow}-c^{\dagger}_{h-x\uparrow}c_{h-y\downarrow}+i(c^{\dagger}_{h-x\downarrow}c_{h-y\uparrow}+c^{\dagger}_{h-x\uparrow}c_{h-y\downarrow})
\nonumber
\\
\!&+&\!c^{\dagger}_{h+y\downarrow}c_{h+x\uparrow}-c^{\dagger}_{h+y\uparrow}c_{h+x\downarrow}+i(c^{\dagger}_{h+y\downarrow}c_{h+x\uparrow}+c^{\dagger}_{h+y\uparrow}c_{h+x\downarrow})
\nonumber
\\
&+& {\rm h.c.}\Big]
\label{hrsc}
\end{eqnarray}
where $h$ is the site occupied by a hole, $h+ x$ ( $h-x$ ) are nearest neighbor sites of $h$ in the $x$ direction (the $-x$ direction); and  $h+y$ ( $h-y$ ) are nearest neighbor sites of $h$ in the $y$ direction (the $-y$ direction). 

In the above Hamiltonian, we have assumed that the Rashba interaction is significant only around the holes with the internal electric field in the direction perpendicular to the CuO$_2$ plane; the electric field is generated by the positive charge of the hole and the compensating charge due to dopant atoms in the charge reservoir layer, and the major component of it is assumed in the direction perpendicular to the CuO$_2$ plane since it is expected that the doped hole is more stable in the position of the CuO$_2$ plane close to the dopant atoms (for example, Sr for La$_{2-x}$Sr$_x$CuO$_4$). The internal electric field created this way will exist even for the cuprates whose parent compounds have a mirror symmetry with respect to the CuO$_2$ plane since the substituted atoms break the local symmetry around the small polaron. However,  the direction of the internal electric field may change either upwards or downwards, locally, with respect to the CuO$_2$ plane. In the present work, we only consider the case where the direction of the internal electric field around the holes is upwards throughout the sample.

\begin{figure}
\begin{center}
\includegraphics[scale=0.6]{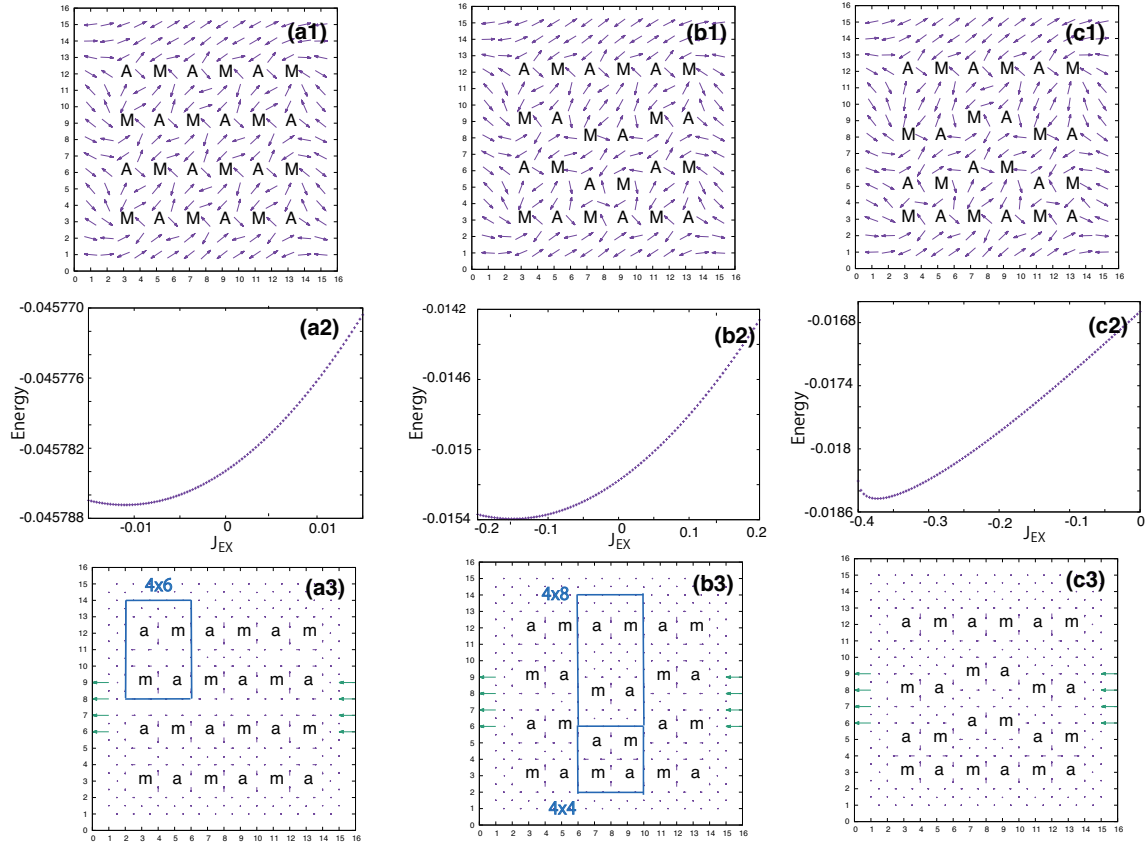}
\caption{(Color online) (a1) Spin-vortices for the system with six $4 \times 6$ spin-vortex quartets in the units of CuO$_2$ plane lattice constant.  `$M$' and `$A$' indicates spin-vortices with winding numbers $+1$ and $-1$, respectively. (a2) The Rashba spin-orbit interaction energy vs the external feeding current $J_{\rm EX}$. The units of energy is $t$, and current is $et / \hbar$, where $t$ is the nearest neighbor hopping integral.
 (a3) The energy minimizing SVILC pattern for (a1) spin-vortex configuration. `$m$' and `$a$' indicates SVILCs with winding numbers $+1$ and $-1$, respectively. Green arrows indicate external current feeding. Each arrows indicating $J_{\rm EX}$ flow-in or flow-out. (b1)-(b3), the same as (a1)-(a3) but for the system with four $4 \times 6$ spin-vortex quartets, one $4 \times 8$ spin-vortex quartet, and one $4 \times 4$ spin-vortex quartet. (c1)-(c3), the same as (a1)-(a3) but for the system with two $4 \times 6$ spin-vortex quartets, two $4 \times 8$ spin-vortex quartets, two one $4 \times 4$ spin-vortex quartets.  $4 \times 6$ spin-vortex quartets is indicated in (a3), and $4\times 8$ and $4\times 4$ spin-vortex quartets are indicated in (b3).}
\label{Fig}
\end{center}
\end{figure}

The results of the Rashba spin-orbit interaction energy dependence of  the external feeding current is depicted in Fig.~\ref{Fig}.
The Hamiltonian used in these calculations is the same one we used in our previous work \cite{Koizumi2017}. The parameter $\lambda$ is taken to be $\lambda=0.01t$, where $t$ is the nearest neighbor hopping integral.
It is remarkable that the energy minima occur at nonzero external current feeding. We can identify this current as supercurrent of superconductivity.  
The energy minima become at zero external-current when the Rashba interaction is absent. Thus, the Rashba interaction is a necessary ingredient for this supercurrent generation.

\section{Possible explanation for the appearance of $8 a$ charge ordered state in the cuprate vortex halo}

As shown in Fig.~\ref{Fig} the the magnitude of supercurrent in (a2), (b2), (c2), depends on the SVILC pattern shown in (a3), (b3), (c3).
This means that each supercurrent has its own SVILC pattern. The lowest energy SVILC pattern among the three is the one in (a3) that is composed of $4 \times 6$ spin-vortex quartets (a spin-vortex quartet is a stable unit composed of four spin-vortices and four SVILCs) \cite{Koizumi2017}.
When some of  $4 \times 6$ spin-vortex quartets are replaced by $4 \times 8$ and $4 \times 4$ spin-vortex quartets, the magnitude of the supercurrent increases as shown in (b2) and (c2) compared with (a2). Actually, it is indicated in our previous work that $4 \times 8$ and $4 \times 4$ spin-vortex quartets may become more favorable in the current flowing situation by the energy gain from the Rashba interaction \cite{Koizumi2017}.

Recently, clear evidence of the appearance of $8a$ charge order was observed in the cuprate vortex halo \cite{Seamus2018}. This may be attributed to the change of SVILC pattern due to the presence of the magnetic-field-screening current (the Meissner current). This SVILC pattern change will involve the transformation from the $4 \times 6$ tiling to $4 \times 8$ tiling. Thus, $8 a$ charge ordered will be enhanced as is observed in the experiment.
It also reduces $6a$ charge order, which is not confirmed, but may be found if the experimental results is reexamined.

\section{Concluding remarks}

Recent developments in condensed matter physics theory point to a necessity for reformulating supercurrent generation mechanism in superconductors. The first impetus comes from a misfit that exists between the experimentally observed ac Josephson effect and the Josephson's prediction \cite{Koizumi2011,HKoizumi2015}. Actually, there is a significant difference in boundary conditions between the Josephson's derivation and the experiment. The Josephson's derivation assumes a simple appearance of a dc voltage across the Josephson junction; however, a dc voltage does not appear by a simple application of a dc voltage; instead, when a dc voltage is applied, a dc Josephson effect takes over, resulting in a zero voltage across the junction. In other words, an electric  power source connected to the Josephson junction actually acts as a current source.

In the experimental situation where a finite voltage exists, there also exist a radiation field and a dc current flow. In this situation, there are two contributions; one from the chemical potential difference between the leads connected to the junction, and the other from the electric field in the non-superconducting region between the two superconductors in the junction. Josephson's derivation takes into account only one of them. The two contributions are equal due to the balance between the voltage and chemical potential difference. By taking into account the two contributions and also the fact that the observed voltage quantization is in the units of ${ {hf} \over {2e}}$, where $f$ is the frequency of the radiation field present, the charge on the carriers should be $q=-e$ \cite{Koizumi2011,HKoizumi2015}. This indicates that the electron pairing is not the true cause of the supercurrent generation, although the pairing energy gap formation temperature is the superconducting transition temperature for many superconductors (but not for cuprates). 

As is described in this work and our previous works, stable current carrying states can be generated when spin-twisting circular motion of electrons occurs. Especially, when the Rashba spin-orbit interaction exists, non-zero current feeding state exhibits an energy minimum; in this situation, persistent current flows through the system without voltage drop. This persistent current can be regarded as supercurrent, thus, the system is in the superconducting state.

Note that it has been shown that spin-twisting circular motion of electrons also occurs in the BCS superconductors if the Rashaba interaction and magnetic field are present \cite{Koizumi2015b}. It occurs due to the fact that the Rashba interaction modifies the electron pairing from the original BCS one to the pairing of spin-twisting cyclotron motion states in the region where the magnetic field is present. Thus, it is suggested that the occurrence of the spin-twisting itinerant electron motion may be the most important ingredient of superconductivity. Such a motion gives rise to the Dirac string with $\pi$ flux. Then, the energy minimizing wave function becomes multi-valued function with respect to electron coordinates and the legitimate single-valued ground state wave function is given as a product of the energy minimizing multi-valued wave function and a U(1) phase factor that compensates the multi-valuedness of the former. 
This phase factor provides a U(1) instanton, explaining the flux quantization and voltage quantization as topological effects \cite{FluxRule}.

\section*{References}




\begin{thebibliography}{99}

\bibitem{Koizumi2011}
Koizumi H. 2011  Spin-vortex Superconductivity. {\em J. Supercond. Nov. Magn.}
  \textbf{24}, 1997.

\bibitem{Hidekata2011}
Hidekata R, Koizumi H. 2011  Spin-vortices and spin-vortex-induced loop
  currents in the pseudogap phase of cuprates. {\em J. Supercond. Nov. Magn.}
  \textbf{24}, 2253.

\bibitem{HKoizumi2013}
Koizumi H, Hidekata R, Okazaki A, Tachiki M. 2014  Persistent current
  generation by the spin-vortex formation in cuprate with the single-valuedness
  constraint on the conduction electron wave functions. {\em J Supercond Nov
  Magn} \textbf{27}, 121.

\bibitem{HKoizumi2014}
Koizumi H, Okazaki A, {Abou Ghantous} M, Tachiki M. 2014  Supercurrent flow
  through the network of spin-vortetices in cuprates. {\em J. Supercond. Nov.
  Magn.} \textbf{27}, 2435.

\bibitem{Koizumi2015b}
Koizumi H, Tachiki M. 2015  Instability of the {BCS} type pairing in magnetic
  field due to the Rashba spin-orbit interaction and supercurrent generation by
  spin-twisting itinerant motion of electrons in BCS superconductors. {\em J
  Supercond Nov Magn} \textbf{28}, 2267.

\bibitem{Bianconi}
Bianconi A, Saini NL, Lanzara A, Missori M, Rossetti T, Oyanagi H, Yamaguchi H,
  Oka K, Ito T. 1996  Determination of the Local Lattice Distortions in the
  {CuO$_2$} Plane of {La$_{1.85}$Sr$_{0.15}$CuO$_4$}. {\em Phys. Rev. Lett.}
  \textbf{76}, 3412.

\bibitem{Miyaki2008}
Miyaki S, Makoshi K, Koizumi H. 2008  Two-Copper-Atom Units Induce a Pseudo
  Jahn-Teller Polaron in Hole-Doped Cuprate Superconductors. {\em J. Phys. Soc.
  Jpn.} \textbf{77}, 034702.

\bibitem{Wu2011}
Wu T, Mayaffre H, Kramer S, Horvatic M, Berthier C, Hardy WN, Liang R, Bonn DA,
  Julien MH. 2011  Magnetic-field-induced charge-stripe order in the
  high-temperature superconductor YBa$_2$Cu$_3$O$_y$. {\em Nature}
  \textbf{477}, 191.

\bibitem{Wu2013}
Wu T, Mayaffre H, Kr{\"a}mer S, Horvati{\'c} M, Berthier C, Kuhns PL, Reyes AP,
  Liang R, Hardy WN, Bonn DA, Julien MH. 2013  Emergence of charge order from
  the vortex state of a high-temperature superconductor. {\em Nature
  Communications} \textbf{4}, 2113.

\bibitem{Wu2015}
Wu T, Mayaffre H, Kr{\"a}mer S, Horvati{\'c} M, Berthier C, Hardy WN, Liang R,
  Bonn DA, Julien MH. 2015  Incipient charge order observed by NMR in the
  normal state of YBa$_2$Cu$_3$O$_y$. {\em Nature Communications} \textbf{6},
  6438.

\bibitem{Ghiringhelli2012}
Ghiringhelli G, Le~Tacon M, Minola M, Blanco-Canosa S, Mazzoli C, Brookes NB,
  De~Luca GM, Frano A, Hawthorn DG, He F, Loew T, Sala MM, Peets DC, Salluzzo
  M, Schierle E, Sutarto R, Sawatzky GA, Weschke E, Keimer B, Braicovich L.
  2012  Long-Range Incommensurate Charge Fluctuations in
  (Y,Nd)Ba$_2$Cu$_3$O$_{6+x}$. {\em Science} \textbf{337}, 821.

\bibitem{Achkar2012}
Achkar AJ, Sutarto R, Mao X, He F, Frano A, Blanco-Canosa S, Le~Tacon M,
  Ghiringhelli G, Braicovich L, Minola M, Moretti~Sala M, Mazzoli C, Liang R,
  Bonn DA, Hardy WN, Keimer B, Sawatzky GA, Hawthorn DG. 2012  Distinct Charge
  Orders in the Planes and Chains of Ortho-III-Ordered
  YBa$_{2}$Cu$_{3}$O$_{6{+}{\delta}}$ Superconductors Identified by Resonant
  Elastic X-ray Scattering. {\em Phys. Rev. Lett.} \textbf{109}, 167001.

\bibitem{Blackburn2013}
Blackburn E, Chang J, H\"ucker M, Holmes AT, Christensen NB, Liang R, Bonn DA,
  Hardy WN, R\"utt U, Gutowski O, Zimmermann Mv, Forgan EM, Hayden SM. 2013
  X-Ray Diffraction Observations of a Charge-Density-Wave Order in
  Superconducting Ortho-II YBa$_{2}$Cu$_{3}$O$_{6.54}$ Single Crystals in Zero
  Magnetic Field. {\em Phys. Rev. Lett.} \textbf{110}, 137004.

\bibitem{Chang2012}
Chang J, Blackburn E, Holmes AT, Christensen NB, Larsen J, Mesot J, Liang R,
  Bonn DA, Hardy WN, Watenphul A, Zimmermann Mv, Forgan EM, Hayden SM. 2012
  Direct observation of competition between superconductivity and charge
  density wave order in YBa$_2$Cu$_3$O$_{6.67}$. {\em Nat Phys} \textbf{8},
  871.

\bibitem{Sebastian2012}
Sebastian SE, Harrison N, Lonzarich GG. 2012  Twoards resolution of the Fermi
  surface in underdoped high-Tc superconductors. {\em Rep. Prog. Phys.}
  \textbf{75}, 102501.

\bibitem{Seamus2018}
Edkins SD, Kostin A, Fujita K, Mackenzie AP, Eisaki H, Uchida SI, Sachdev S,
  Lawler MJ, Eun-Ah~Kim JCSD, Hamidian MH. 2018  Magnetic-field Induced Pair
  Density Wave State in the Cuprate Vortex Halo. {\em arXiv:1802.04673
  [cond-mat.supr-con]}.

\bibitem{Koizumi2017}
Morisaki T, Wakaura H, Koizumi H. 2017  Effect of Rashba Spin--Orbit
  Interaction on the Stability of Spin-Vortex-Induced Loop Current in
  Hole-Doped Cuprate Superconductors. {\em J. Phys. Soc. Jpn.} \textbf{86},
  104710.

\bibitem{Rashba2013}
Riera JA. 2013  Spin polarization in the Hubbard model with Rashba spin-orbit
  coupling on a ladder. {\em Phys. Rev. B} \textbf{88}, 045102.

\bibitem{HKoizumi2015}
Koizumi H, Tachiki M. 2015  Supercurrent Generation by Spin-twisting Itinerant
  Motion of Electrons: Re-derivation of the ac Josephson Effect Including the
  Current Flow Through the Leads Connected to Josephson Junction. {\em J
  Supercond Nov Magn} \textbf{28}, 61.

\bibitem{FluxRule}
Koizumi H. 2017  Flux rule, {U}(1) instanton, and superconductivity. {\em J.
  Supercond. Nov. Magn.} \textbf{30}, 3345.

\end{thebibliography}

\end{document}